# Towards an AI-Enhanced Cyber Threat Intelligence Processing Pipeline

Lampis Alevizos
*University of Central Lancashire (UCLan) -*
*School of Engineering and Computer Science*
Preston, United Kingdom
lampis@redisni.org

Martijn Dekker
*University of Amsterdam (UvA) - Amsterdam Business School*
*Faculty of Economics and Business*
Amsterdam, The Netherlands
m.dekker4@uva.nl

*Abstract* — Cyber threats continue to evolve in complexity, thereby traditional Cyber Threat Intelligence (CTI) methods struggle to keep pace. AI offers a potential solution, automating and enhancing various tasks, from data ingestion to resilience verification. This paper explores the potential of integrating Artificial Intelligence (AI) into CTI. We provide a blueprint of an AI-enhanced CTI processing pipeline, and detail its components and functionalities. The pipeline highlights the collaboration of AI and human expertise, which is necessary to produce timely and high-fidelity cyber threat intelligence. We also explore the automated generation of mitigation recommendations, harnessing AI's capabilities to provide real-time, contextual, and predictive insights. However, the integration of AI into CTI is not without challenges. Thereby, we discuss ethical dilemmas, potential biases, and the imperative for transparency in AI-driven decisions. We address the need for data privacy, consent mechanisms, and the potential misuse of technology. Moreover, we highlights the importance of addressing biases both during CTI analysis and AI models warranting their transparency and interpretability. Lastly, our work points out future research directions such as the exploration of advanced AI models to augment cyberdefences, and the human-AI collaboration optimization. Ultimately, the fusion of AI with CTI appears to hold significant potential in cybersecurity domain.

*Index Terms*—artificial intelligence, cyber threat intelligence, cyber resilience, ethical considerations, CTI and AI biases.

## I. INTRODUCTION & MOTIVATION

Cyber threats continuously grow in complexity and frequency, therefore the ability to rapidly process and act upon cyber threat intelligence (CTI) can mean the difference between a mitigated threat and a breach. CTI, as defined by the National Institute of Standards and Technology (NIST), includes information that allows organizations to understand the latest threats and to proactively defend themselves against them [1]. However, the vast volume of CTI, coupled with its dynamic nature, poses significant challenges for timely processing and action.

Traditional CTI processing methodologies involve manual efforts, where analysts examine large amounts of data, attempting to recognise patterns, validate intelligence, and recommend actions [2]. Namely, analysts are trying to produce actionable and valuable intelligence by contextualizing information. This manual approach, while valuable, is increasingly becoming unsustainable given the scale and speed of modern cyber threats. The need for automation and enhanced analytical capabilities thereby becomes evident.

AI with its ability in handling large datasets and its capability to learn and adapt, offers a promising direction to augment the CTI processing pipeline. Preliminary research, such as the works of Buczak and Guven [3], have already highlighted the potential of AI in cybersecurity, particularly in areas like anomaly detection and malware classification. However, the integration of AI into the CTI processing pipeline, especially in a manner that highlights the collaboration with human expertise, remains an area of research.

This paper seeks to bridge this gap, presenting a comprehensive approach to harnessing AI for CTI processing. Our focus goes beyond automation, creating a collaborative framework where AI and human analysts work together, to produce rapid, accurate, and actionable CTI. By streamlining this pipeline, we aim to reduce the time from intelligence ingestion to the implementation of mitigating measures and subsequent resilience verification. We believe that combining AI with CTI offers a proactive and adaptable cybersecurity approach, rather than a reactive one. Our goal is to connect AI's capabilities with cybersecurity requirements, advancing future innovations in the field. The contributions of this paper can be summarized as follows.

**(1) Blueprint of the AI-enhanced CTI processing pipeline:** we present a comprehensive framework that integrates AI techniques at various stages of a threat-informed defence, starting with CTI data ingestion down to resilience verification. We detail the components and functionalities as well as highlighting the imperative collaboration between AI and human expertise.

**(2) Innovation in real-time and predictive threat mitigation:** Our research pioneers the use of AI for generating real-time, contextual, and predictive mitigation strategies. We explore the application of advanced AI algorithms that can swiftly analyse CTI data and suggest security measures, thereby enhancing organizational responsiveness to cyber threats.

**(3) Ethical and bias considerations:** we performed a thorough examination of the ethical implications of using AI in CTI and strategies to address potential biases in both CTI domain and AI models. We also propose methods to ensure unbiased and transparent AI-driven insights.

**(4) Introduction of a Cyber resilience index:** we propose a novel cyber resilience index that serves as a barometer for an organization's defensive capabilities against cyber threats. Analogous to financial market indices, this metric offers a quick overview of an organization's cyber health, informing strategic defence decisions.

**(5) Challenges in AI-Driven CTI:** we critically discuss the hurdles in embedding AI into CTI, starting with ethical dilemmas, data bias, and the need for transparency in AI-driven decisions, presenting a roadmap for addressing these issues.

**(6) Future research directions:** we provide future research directions emerging from our findings, thus, underlining areas of potential growth and innovation.

## II. BACKGROUND & LITERATURE REVIEW

The convergence of CTI and AI is a relatively emerging field, but one that has gained significant attention due to the potential benefits it promises. CTI has evolved from basic threat feeds to sophisticated intelligence platforms that provide contextual information about threats [4]. The primary goal of a mature CTI capability should be to offer actionable and valuable insights that can guide defensive measures, essentially extracting the right signals throughout the vast "noise" within the cyber landscape [5]. The works of Chen et al. [6] provide a comprehensive overview of the CTI landscape, highlighting the challenges associated with intelligence validation and relevance determination.

The application of AI in cybersecurity is not completely new. Machine learning models have been employed for tasks like spam detection and network intrusion detection for years as detailed in the work of Sarker et al. [7]. However, the integration of AI with CTI is a more recent endeavour and lacks research output. The potential of AI to process vast amounts of data rapidly makes it a natural fit for CTI processing. for instance, Ring et al. [8], explored the use of AI for threat hunting, highlighting the potential to uncover hidden threats in vast datasets.

Despite the advancements, several challenges persist in CTI processing. The dynamic nature of the cyber threat landscape, coupled with the large volume of data, often leads to information overload [9]. Additionally, false positives, outdated intelligence, and the lack of context can hamper the effectiveness of CTI. Sauerwein et al. [10] researched these challenges, offering insights into potential mitigation strategies.

The collaboration between AI and human expertise is also an important topic of considerable interest. While AI excels at processing large datasets, human intuition and expertise remain irreplaceable for nuanced threat analysis [11]. The challenge lies in creating a framework or pipeline where AI augments human capabilities without overwhelming them with data. Brundage et al. [12] discussed the potential pitfalls and best practices for AI-human collaboration in decision-making contexts. In our work, we aim to bridge this gap by establishing strong collaborative bonds between the CTI analyst and AI, where both complement each other's strengths and counter each other's weaknesses.

Varma et al. [13] work primarily targets small and medium-sized enterprises (SMEs), providing them with a roadmap to integrate AI into their CTI processes. Although it offers valuable insights for SMEs, its scope is limited when considering larger organizations or more complex cybersecurity infrastructures. The researchers used AI only to enhance CTI, rather than a comprehensive solution for cybersecurity. Our paper on the other hand, presents a comprehensive AI-enhanced CTI processing pipeline adaptable to organizations of any size. Additionally, the roadmap in the referenced work can be seen as a practical application, while our proposed pipeline offers a broader perspective, including all stages of CTI processing, making it more universally applicable while also allowing organizations to select the components that best suit their needs. Suryotrisongko et al. [14] underlined the importance of trust in CTI sharing, advocating for the blending of explainable AI (XAI) with open-source intelligence (OSINT). While it underlines the significance of transparency in AI-driven CTI, it primarily focuses on botnet detection, which is a specific subset of the broader CTI landscape. Our focus on transparency and interpretability in AI-driven insights aligns with the principles of XAI. However, our goal is to provide a more holistic view of the CTI landscape, addressing various challenges and stages of CTI processing beyond just botnet detection. Ranade et al. [15] studied a niche but crucial challenge in CTI, namely the generation of fake CTI descriptions using advanced AI models. Although it highlights an emerging threat in the CTI domain, its primary focus is on the generation aspect rather than mitigation or validation. In our work we highlight the importance of validation and relevance determination in CTI. The challenge of fake CTI generation further highlights the need for robust validation mechanisms, which our paper addresses in detail, offering solutions and strategies to counter such threats.

Moraliyag et al. [16] proposed a proactive approach to CTI by classifying onion services based on content. Although it provides valuable insights into dark web intelligence, its primary focus remains on classification techniques, potentially overlooking other crucial aspects of CTI processing. Our intelligence ingestion phase in the proposed pipeline can benefit from such classification techniques. Nonetheless, we propose a more comprehensive view, detailing various stages of CTI processing and addressing challenges beyond just classification. Mitra et al. [17] research focuses on enhancing cybersecurity knowledge graphs with intelligence provenance, which is a novel approach to combat fake CTI. Nevertheless, relying solely on provenance might not address all challenges associated with fake CTI, especially when considering sophisticated adversarial attacks. The integration of provenance information aligns with our pipeline's intelligence ingestion and collaborative analysis phases. However, in this work we provide a multi-faceted approach to CTI validation, which enables a more robust defence against fake intelligence. Mittal et al. [18] discussed the potential of AI in CTI, displaying its role in uncovering hidden threats. Whilst it offers valuable insights, it does not provide a detailed roadmap or framework for integrating AI into CTI processing. Our work builds on this premise but goes a step further by detailing how AI can be systematically integrated into various stages of CTI processing, offering a more structured approach.

## III. THE AI-ENHANCED CTI PROCESSING PIPELINE

The fusion of AI capabilities with CTI processing has the potential to significantly enhance the speed, accuracy, and efficiency of threat intelligence operations. This section outlines a structured pipeline that integrates AI at various stages of CTI processing to enable the abovementioned attributes. We draw figure 1 to visualize the individual components comprising the AI-enhanced CTI processing pipeline as a blueprint.

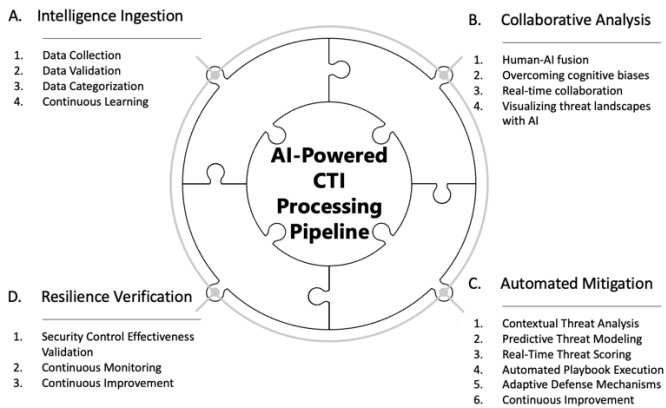

*Figure 1 - AI-enhanced CTI processing pipeline blueprint.*

The central theme of the mind map is represented by the largest circle (blue) labelled "AI-enhanced CTI Processing Pipeline." Radiating outward from this central theme are four main components, namely:

A. **Intelligence ingestion** focuses on the initial stages of data collection, data validation, and data categorization using AI.
B. **Collaborative analysis** focuses on the collaborative analysis between human intelligence and artificial intelligence. We detail the concept of human-AI fusion, ways of overcoming cognitive biases, real-time collaboration, and visualizing threat landscapes with AI.
C. **Automated mitigation** focuses on analysing threats in context using AI. It contains predictive threat modelling, real-time threat scoring, automated playbook execution, adaptive defence mechanisms, a feedback loop for continuous improvement.
D. **Resilience verification** comprises of the proactive approach to security by simulating cyber-attacks. The focus is on continuous monitoring and continuous improvement led by the AI, ultimately leading into a single cyber resilience metric, the cyber resilience index.

Each of these main components further breaks down into subcomponents, that we detail in the corresponding sections.

### A. Intelligence Ingestion

The initial phase of the CTI processing pipeline is the ingestion of raw information or intelligence data. This means collecting, validating, and categorizing vast amounts of data from various sources, such as threat feeds, logs, and other intelligence or information repositories. The threat landscape is dynamic with new threats emerging regularly, thereby it is imperative for a CTI ingestion process empowered by AI, to continuously learn from new data and adapt to the evolving threat environment. Figure 2 outlines the intelligence ingestion steps.

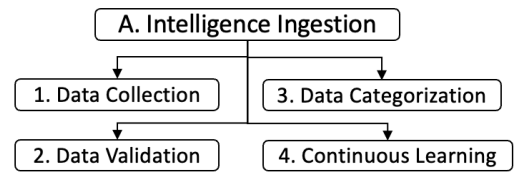

*Figure 2 - Intelligence ingestion steps.*

#### 1) Data Collection

Given the vast amount of CTI data that needs to be collected, manual collection, although feasible, is extremely time-consuming and the added value of such exercise is highly dependent on the analysts experience and expertise alone. AI-driven tools orchestrate multiple APIs (Application Programming Interfaces) and therefore automate the collection process, thus, data is gathered in real-time and from a wide range of sources, such as, Open-source intelligence (OSINT), Human intelligence (HUMINT), dark web monitoring, commercial threat feeds, internal organization threat data, vendor reports, social media monitoring, threat intelligence platforms (TIPs), and industry-specific threat reports.

#### 2) Data Validation

Not all collected data is relevant. AI algorithms can quickly examine the data, discarding irrelevant information and highlighting potential threats using decision trees, as shown by Kotsiantis [19]. Then they cross-reference data from multiple sources to verify its accuracy using graph analytics to map relationships between sources. For instance, if two independent sources report the same threat, it is more likely to be credible. However, the challenge is not just about collecting data but producing meaningful and actionable intelligence from the overwhelming "noise". Transforming information into actionable, valuable intelligence should be the goal. The total volume of data, combined with its dynamic nature, makes manual validation and analysis a challenging task. This is where AI plays a transformative role and can be implemented based on the following elements.

**Signal extraction** using convolutional neural networks (CNNs) provide proven pattern recognition within large datasets, enabling them to detect indirect signs of malicious acticity that may signify cyber threats. This capability allows AI to effectively extract the "signal" – meaningful, actionable intelligence – from the "noise" – irrelevant or redundant information. Goodfellow et al. [20] provided the foundational theory on the capabilities of neural networks in pattern recognition and anomaly detection. TensorFlow[1] serves as the practical implementation being a popular open-source machine learning framework that offers long short-term memory (LSTM) autoencoders which can be used for anomaly detection in time-series data [21].

**Cross-referencing and correlation** utilizing AI tools automatically from multiple data sources, enhancing its validation process. For instance, if two independent sources report the same threat, the AI assigns a higher credibility score to that piece of intelligence. Advanced algorithms can also correlate seemingly unrelated pieces of information, uncovering hidden threats or tactics used by adversaries. Prime theoretical example provided by Landwehr et al. [22] introduced the logistic model trees, which can be used for correlating and cross-referencing data from multiple sources. Elastic Stack[2] (Elasticsearch, Logstash, Kibana) is the practical implementation widely used in cybersecurity for its capabilities in data ingestion, indexing, and visualization. It can correlate logs from various sources to provide a unified view of events.

**Transforming information to intelligence**: AI bridges this gap by analysing the context in which data is generated, determining its relevance to the organization's threat landscape, and providing actionable recommendations based on the analysed intelligence [23]. However, it is important to note that the effectiveness of the AI is depending upon the accuracy and quality of the underlying data, e.g., data within a Configuration Management Database (CMDB) of a company. The quality and integrity of data sources is therefore crucial, as they directly impact the reliability of the intelligence generated.

**Addressing Human Limitations**: Traditional CTI relies on human analysts who, despite their expertise, have limitations in terms of processing capacity and speed. AI complements human analysts by managing vast datasets, ensuring that no potential threat goes unnoticed [3]. This collaboration between human intuition and AI's computational aptitude provides for comprehensive threat intelligence.

A feedback mechanism where false positives or irrelevant data flagged by AI are reviewed and fed back into the system for continuous learning and improvement. This iterative process improves AI model's accuracy over time. Figure 3 demonstrates the process of AI-driven validation, and the transformation of raw data into actionable intelligence.

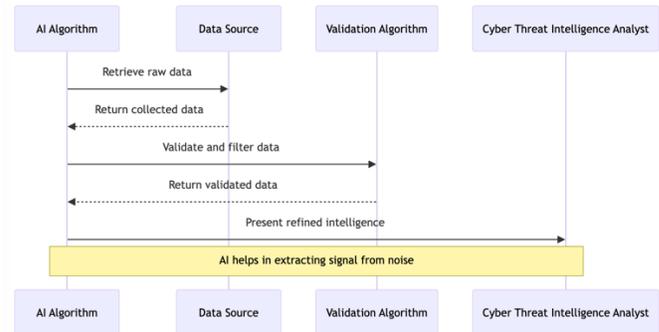

Figure 3 - Extracting the signal from the noise using AI.

The AI algorithm retrieves raw data from the data source. The collected data is returned to the AI algorithm. Next, the AI algorithm sends this data to the validation algorithm to validate and filter out irrelevant or noisy data. The validated data is then returned to the AI algorithm. Finally, the refined intelligence is presented to the CTI analyst.

*3) Data Categorization*

Effective CTI requires the ingested data to be categorized into meaningful segments that can guide subsequent analysis and action from the human-AI collaboration. Latent Dirichlet Allocation (LDA) algorithm can be used to identify the underlying themes or topics within large text datasets, helping to categorize content by subject matter [24]. For instance, threat actors (TAs), threat events, TTPs (Tactics, Techniques, and Procedures), Indicators of Compromise (IoCs), goals and motivations of TAs, and geopolitical trends. Next, Schonlau et al showed how to use random forest [25] while Sarker showed how to use neural networks to classify data into predefined categories based on training datasets [26] where the classification criteria are already known. Thus, using either method AI takes unstructured data and organize it into meaningful segments, ready for further analysis and action in the CTI pipeline.

*4) Continuous learning*

The cyber threat landscape is not static; it evolves continuously with new vulnerabilities, TTPs, and threat actors emerging regularly. For an AI-enhanced CTI pipeline to remain effective, it must adapt to these changes. To successfully enable the AI enhanced CTI pipeline to continuous, learn, several methods can be used.

**Machine learning** (ML) for continuous relevance in CTI utilizing the online learning algorithm [27]. This algorithm incrementally updates parameters in response to each new data point, thus, providing adaptability to emerging threats without full retraining. This approach keeps the predictive model current with the evolving threat landscape.

**Adversarial machine learning (AML)** to anticipate potential evasion techniques that adversaries might employ. Red teams can either perform traditional attack simulations or use AML to simulate advanced evasion tactics. This will result in collecting data on novel attack vectors and improving defences before they are exploited in the wild. Red teams should create adversarial examples led by cyber threat intelligence to assess the organization's defences. Any successful evasion that is logged and analysed, is in turn used to improve data collection mechanisms. As the AI system processes new threat intelligence and interacts with human analysts, it will inevitably encounter false positives or misclassifications. Therefore, incorporating feedback from these interactions, will allow the system to refine its algorithms, reducing errors over time.

---

[1] https://www.tensorflow.org/tutorials/generative/autoencoder

[2] https://www.elastic.co/

**Defensive distillation** should be used to make machine learning models more robust against adversarial attacks [28]. Therefore, the data being collected will not be polluted by adversarial noise. Leveraging this technique to train the model on a "softened" version of the data, where the output probabilities are smoothed or "distilled" will make it harder for adversaries to find the precise distresses needed to deceive the model, thus, cleaner data collection.

**Incorporating broader context** for the AI system should continuously learn and incorporate insights from broader geopolitical, technological, and socio-economic contexts, enhancing its threat predictions. For instance, if an innovative technology becomes widespread (e.g. blockchain), the AI system will prioritize threats targeting that technology.

**Active learning is** a specialized form of machine learning where the model actively queries the human analyst for inputs on specific predictions [29]. For instance, if the AI system encounters a piece of data, it is uncertain about, it seeks confirmation from a human expert. Over time, these interactions reduce the system's uncertainty and improve its accuracy.

### B. Collaborative Analysis

Traditional analysis methods relying heavily on human expertise are oftentimes slow, potentially biased, and prone to errors. This is where artificial intelligence, and more specifically machine learning, can play a crucial role [30]. Mishra's work [31] showed that Gradient Boosting Machines (GBMs), trained on historical threat data, provide insights into potential threats, their patterns, and possible implications. Human analysts collaborate with these AI insights, leveraging their expertise to understand the nuances and context behind each threat. Figure 4 outlines the collaborative analysis steps.

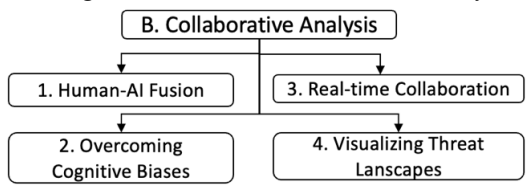

*Figure 4 - Collaborative analysis steps.*

---

[3] https://www.cisco.com/c/en/us/products/security/threat-grid/index.html

#### 1) The human-AI fusion

The collaboration between human analysts and AI is not about replacing one with the other but about amplifying the strengths and mitigating the weaknesses of both. A summary of this phase is visualized in Figure 5.

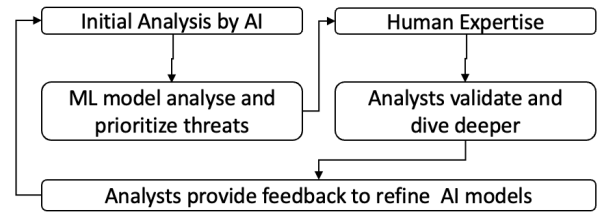

*Figure 5 - The Human AI fusion.*

**Initial analysis by AI provides** speed and scale. AI processes vast amounts of data at speeds incomprehensible to humans. For instance, Desmond et al. [32] showed that a model trained on extensive datasets can analyse millions of logs within minutes to detect anomalies as opposed to humans. Chen's work [33] showed that AI can also provide pattern recognition, through a deep learning algorithm which is able to recognize patterns in data. Goodfellow et al. [20] proposed neural networks that can identify patterns associated with malware traffic in network logs, even if the malware is a zero-day variant. AI offers fast prioritization, in addition. Based on the detected patterns and historical data, AI can prioritize threats, therefore the most imminent and dangerous threats are addressed first. Existing AI-driven tools like Cisco's Threat Grid[3] analyse millions of samples daily, providing automated threat scores based on the potential impact and prevalence of detected threats [34].

**Human expertise provides** broader contextual understanding. Although AI can recognize patterns, human analysts understand the broader context. For instance, in the SolarWinds attack[4] while AI tools detected anomalies, it was the human analysts who pieced together the broader campaign, understanding the implications and the actors behind it. Human analysts bring intuition and experience. Analysts, with years of professional experience, can subjectively understand when something does not seem right, even if it passes AI checks. Their expertise allows them to focus into complex threats specific to the IT landscape, forming hypothesis, and therefore uncovering potential hidden connections. Lastly, human analysts play a crucial role in validating AI findings. Although AI might flag a potential phishing email based on certain patterns, a human analyst can validate it by considering the sender's context, the email's content, and other information. Analysts interacting with the AI-enhanced pipeline, must provide feedback for refining the AI model, thus its predictions become more accurate over time. However, a challenge with AI in cybersecurity is the potential for false positives [10]. Human analysts should flag something as a false positive, eventually,

---

[4] https://www.wired.com/story/the-untold-story-of-solarwinds-the-boldest-supply-chain-hack-ever/

the AI learns from it, thereby reducing similar false alarms in the future.

*2) Overcoming Cognitive Biases*

Cognitive biases are systematic patterns of deviation from norm or rationality in judgment, thus, leading analysts to create their own subjective reality from their perception of the input [35]. Such biases can significantly impact the decisions of CTI analysts, thereby potentially leading to disregarded threats or data misinterpretations.

One of the major ransomware[5] related attacks happened in 2017. The aftermath of this case, was that many organizations focused heavily on protecting against similar ransomware threats. Although this is a valid concern, an overemphasis on one type of threat due to its recent occurrence (availability heuristic) can lead to neglecting other potential threats. The AI-driven CTI pipeline would provide for a balanced focus on all relevant threats, not just those that are currently in the spotlight. The integration of AI into the CTI process offers a unique opportunity to counteract these biases, ultimately allowing for a more objective and comprehensive analysis.

**CTI analysists may be subject to the following biases:**
- **Confirmation bias,** the analyst might prioritize data that aligns with their existing threat models, potentially overlooking new or unexpected threats. The analysts perspective is unintentional influencing the collection, analysis, and interpretation of CTI data. This bias (also known as observer bias) can lead analysts to favor information that confirms their presumptions or to overlook data that contradicts their beliefs, hence impacting the accuracy and objectivity of their analysis
- **Availability heuristic,** the analyst might give undue weight to a recent high-profile cyberattack, neglecting other potential threats.
- **Anchoring bias,** the analyst relies too heavily on the first piece of information encountered (the "anchor") when making decisions. Oftentimes CTI analysts anchor their analysis directly on initial findings, therefore missing the broader threat landscape.
- **Status quo bias,** a preference for the current situation, resisting change, leading into an over-reliance on established threat models and an inability to adapt to the evolving cyber threat landscape.

**AI's role in mitigating biases:**
- AI algorithms are inherently decoupled from emotions and preconceived notions, thereby provide **an objective analysis of data**. They treat each piece of information based on its merit and relevance, not on any external influence or bias [36].
- AI models apply **consistent criteria** when analysing data, thus using the same standards across, contrary to the human analyst, who might unconsciously alter their criteria based on biases [37].
- AI analyses data for decisions **based on comprehensive data** rather than anecdotal evidence or recent events [37].

Conclusively, to successfully overcome cognitive biases, the goal is not to replace human analysts with AI but to have them work together. AI provides objective analysis, empowered by human analysts bringing contextual understanding and intuition. By working together, they can counteract the biases inherent in both human judgment and AI models, therefore leading to a more balanced and comprehensive threat analysis.

In exploring how humans and AI work together to counteract biases and optimise collaboration, Dell'Acqua et al. [38] highlighted two main ways: 'Cyborg' and 'Centaur.' The 'Cyborg' mode mixes human and AI efforts closely, using AI for its fast processing and humans for their deep understanding and moral judgment. On the other hand, the 'Centaur' mode is about humans and AI working side by side, with each taking on tasks that suit their strengths accordingly. This division helps make the most of both AI's data handling abilities and human creativity and ethical insights. These dimensions can serve as guardrails and show how combining human and AI strengths can enhance strategic decision making (centaur) and improve efficiency and accuracy (cyborg).

*3) Real-time Collaboration*

Real-time machine learning models, bring a change in thinking in how threat intelligence is processed and acted upon. Consider a zero-day vulnerability that has just been discovered. Traditional threat intelligence systems might take hours, if not days, to update their databases and provide recommendations. However, a real-time AI-driven system can pick up discussions about this vulnerability from sources like social media, forums, commercial tools, TIPs, or dark web marketplaces within minutes. It can then assess the potential impact of this vulnerability, generate alerts for human analysts, and even recommend immediate countermeasures [39]. The successful real-time collaboration between the CTI analyst and AI, is based on the following elements.

**Dynamic data ingestion**, as cyber threat data is generated by any of the above-described sources continuously, it is imperative to have a system that can ingest this data in real-time. AI-driven models, especially those built on streaming data platforms, can process data as it flows in, without waiting for batch updates [40].

**Instantaneous analysis** once the data is ingested. Real-time machine learning models analyse it instantaneously [41]. This means that as soon as a new threat indicator is detected, the AI-enhanced CTI pipeline can assess its severity, potential impact, and relevance.

**Real-time alerts** based on the instantaneous analysis; the AI enabled CTI pipeline generates real-time alerts for human analysts. These alerts can be prioritized based on potential impact, and thereby the analysts can focus on the most pressing threats first.

---
[5] https://www.csoonline.com/article/563017/wannacry-explained-a-perfect-ransomware-storm.html

**Human-AI interaction**; the real-time collaboration should not be just one-way. Human analysts, upon receiving alerts, can interact with the AI enhanced CTI pipeline, asking follow-up questions or clarifications.

**Adaptive learning** is one of the differentiating factors of real-time machine learning models is their ability to learn on-the-fly. As new data is processed, the model can update its understanding, hence its predictions and recommendations are always based on the latest threat intelligence [42].

*4) Visualizing threat landscapes with AI*

The ability to visualize and quickly comprehend threat models is paramount. As an example, one could think of a scenario where a CTI analyst comes across an image detailing the flow of a sophisticated malware attack. Instead of spending hours or even days deciphering the image, the analyst can use an AI tool to quickly understand the malware's propagation, its potential targets, and its behaviour. The AI tool can also cross-reference the malware's signature with a database, providing insights into its origin, past variants, and potential countermeasures [43]. There are two prime examples where AI can supercharge the collaborative analysis and empower the CTI analyst:

**(i) Automated threat modelling:**
Traditional threat modelling is a time-consuming exercise, requiring analysts to manually map out threats, vulnerabilities, and potential attack vectors. Moreover, CTI analysts and relevant stakeholders may lack the technical knowledge to perform threat modelling. Akhtar et al. [44] showed how AI automates this process, rapidly creating threat models based on the available data. Moreover, as new threat intelligence is ingested, AI can dynamically update the threat model, thus always reflecting the current threat landscape. Lastly, AI algorithms and especially deep learning models can be used to identify difficult correlations and potential threats that might be overlooked in manual analysis [45].

**(ii) Image recognition and explanation:**
CTI analysts can drop images depicting complex IT landscapes or complex threat actor flows into an AI-powered tool. Iqbal et al. showed how the AI can instantly analyse the image, recognizing various components, connections, and potential vulnerabilities [46]. Furthermore in the same work, it was proven that AI can go beyond simple recognition. Advanced AI models can provide contextual explanations for the elements in the image [46]. For instance, if an analyst drops an image of a network topology, the AI can identify servers, firewalls, potential choke points, and even suggest potential attack vectors. based on the layout. Furthermore, by cross-referencing the elements in the image with historical threat data, AI can provide insights into past vulnerabilities, attacks, or breaches associated with similar setups [45].

### C. Automated mitigation

Based on the combined intelligence from AI and human analysis, the pipeline integrates with organizational tools like Configuration Management Databases (CMDBs) or taps into IT infrastructure data to understand the environment and adapt recommendations. The recommendations range from technical solutions, like updating firewall rules or patching vulnerabilities, to strategic actions, such as user awareness campaigns or policy changes. In this section we outline how AI can be harnessed to provide automated mitigation recommendations based on analysed intelligence, and to visualize the integrated approach of AI-driven mitigation

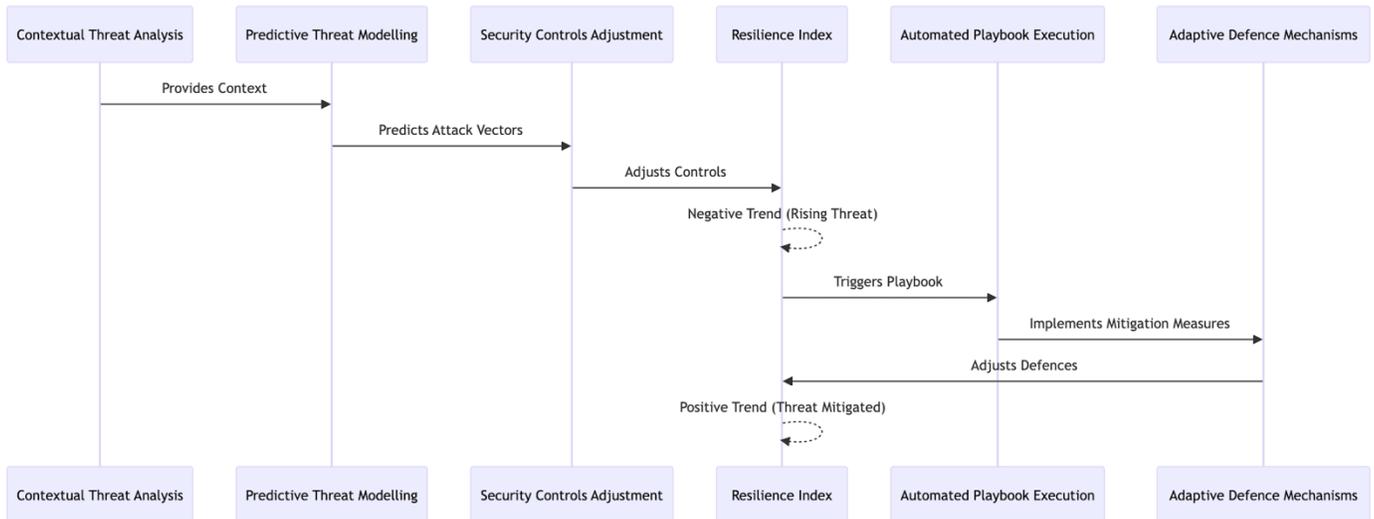

*Figure 6 – AI Driven CTI pipeline security control steering.*

*1) Contextual threat analysis*

Before the AI-enhanced CTI pipeline recommends any mitigation strategy, it is crucial to understand the context of the threat against the operating IT landscape. Therefore, analysis ot the threat in relation to the organization's infrastructure, assets, and previous incidents is necessary. This is achieved using

Natural Language Processing (NLP) to extract contextual information from threat intelligence reports [47].

Suppose an organization receives a threat intelligence report about a new ransomware strain targeting financial institutions. Using NLP, the AI system extracts keywords like "ransomware," "financial institutions," and cross-reference this with the organization's IT and security landscape to determine the relevance and potential impact, thereby adjusting the relevant security controls correspondingly.

*2) Predictive threat modelling*

AI trained with current intelligence and historical data can predict the likely progression of a threat based on patterns observed in past incidents [48]. As a result, we can pre-emptively strengthen the defences in vulnerable areas. For example, if historical data indicates that every time there is a spike in traffic from a particular region, a DDoS attack follows, the AI can predict a potential DDoS attack when it observes a similar traffic pattern in the future and therefore proactively adjust relevant security controls accordingly.

*3) Real-time threat scoring*

Not all threats have the same level of severity or relevance to an organization. AI provides a real-time threat score based on specific IT landscape and organization information, helping prioritize mitigation efforts. Next, it can score threats based on factors like potential impact, exploitability, and the organization's vulnerability faster [49]. As a result, the most critical threats are addressed first. For instance, an organization might receive thousands of alerts daily. An AI system can score a detected phishing attempt as "high" if it is linked to a known APT group while a generic malware detection might be scored as "medium".

*4) Automated playbook execution*

For known threats or attack patterns, AI automatically executes predefined mitigation playbooks, reducing the response time subject to integration with threat intelligence systems and Security Orchestration, Automation, and Response (SOAR) platforms. Upon detecting a recognized threat pattern, the pipeline triggers the corresponding playbook for immediate action. For example, if the AI pipeline detects patterns consistent with the "Emotet" malware, it can trigger a predefined playbook that isolates affected systems, blocks associated IPs, and sends notifications to the incident response team. Although the AI-enhanced CTI pipeline speeds up threat response by running preset mitigation strategies, it is imperative to involve humans in key decision-making to handle high-risk situations. For instance, cybersecurity experts should review and authorize actions chosen by the AI in high-risk scenarios, combining the speed of AI with human insight to minimize risks of automated responses.

*5) Adaptive Defence Mechanisms*

AI dynamically adjusts defences based on ongoing threat analysis using reinforcement learning (RL) models that can adapt security configurations in real-time [50]. For instance, if the AI pipeline detects increased traffic from a specific IP range associated with malicious activities, it can dynamically adjust firewall rules to block or throttle that traffic. Or if the AI pipeline observes that every Friday evening there is an attempt to exfiltrate data, it can dynamically adjust egress firewall rules during that time to add an additional layer of scrutiny.

*6) Continuous improvement*

After implementing mitigation measures, it is essential to assess their effectiveness and refine strategies accordingly. For instance, after blocking a suspected malicious IP, the AI system can monitor for any subsequent attempts or changes in attack patterns from that IP, refining its threat profile over time.

### D. Resilience Verification

Simulation or emulation of attack scenarios to assess the resilience of the implemented measures is imperative at this stage. Security control effectiveness evaluation measures the resilience of organizations against potential cyber threats, eventually producing an alternative to a stock market index, but for cybersecurity. Much like financial indices provide traders with a snapshot of the market's health or trend, e.g., S&P 500 index, a cyber resilience index can offer decision-makers a quick overview of their organization's cybersecurity posture. This index, updated by the AI in real-time, can serve as a barometer for an organization's cyber health, and therefore used by decision makers to steer their defences and resources accordingly. As a result, organizations are not just reactive, but pro-active in their cybersecurity approach.

To achieve this, we define the following three steps to resilience verification leading into the formation of a cyber resilience index: (1) automated penetration testing (2) continuous monitoring, and (3) continuous improvement illustrated in figure 7.

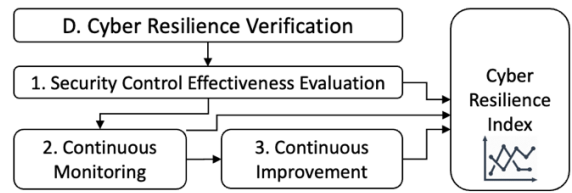

Figure 7 - Resilience verification steps.

*1) Security control effectiveness evaluation*

The AI provides a governance layer on processes or tools to simulate or emulate cyber-attacks on a system to identify vulnerabilities and assess the effectiveness of the implemented mitigation measures. As a result, it simulates and emulates advanced persistent threats (APTs) to assess how well a system can withstand prolonged, targeted attacks. Moreover, since the AI governs the CTI pipeline, it can adapt the strategies of APTs based on received CTI and against the system's responses, mimicking the behaviour of real-world adversaries [51]. Therefore, providing factual security control effectiveness validation rather than checklist-based theoretical assessments.

*2) Continuous monitoring*

The AI continuously monitors network traffic, system logs, and other relevant data sources from the IT landscape it is deployed. As a result, it provides real-time detection of suspicious activity that might indicate a breach or vulnerability exploitation. It can also be trained on historical network traffic data to recognize patterns related to known cyber threats, thus serving as an AI-power intrusion detection system (IDS). Once deployed, it can monitor network traffic in real-time, flagging any deviations from the norm for further investigation.

*3) Continuous Improvement*

One of the key advantages of integrating AI into resilience verification is the ability to continuously self-improve. As an illustration, the AI has the ability to observe and evaluate the ecosystem in which it operates, and can gain knowledge from any newly identified threats or weaknesses, improving its algorithms for the next evaluations. For instance, when an intrusion detection system identifies a new form of malware on the network, it can adjust the AI algorithms to identify this threat in the future.This continuous learning helps the pipeline to be updated with the latest threat intelligence, coupled with the latest data from the IT landscape it is deployed on.

## IV. CHALLENGES AND CONSIDERATIONS

Although an AI-enhanced CTI processing pipeline may offer significant capabilities, it comes with several challenges and considerations that we discuss in this section. Given that we are in the nascent stages of building AI for corporate use it is imperative for organizations starting to work on AI technologies, and especially within CTI, should follow guidelines set by NIST AI RMF 1.0 [52], the EU AI Act [53] [54], and ISO 5338 [55]. Adhering to these standards will help organizations develop AI systems that are effective, efficient, ethically responsible, and compliant with global regulations.

*A. Ethical Consideration in AI-enhanced CTI Processing*

*1) Data Privacy and Confidentiality*

AI-enhanced CTI processing requires access to vast amounts of data, some of which may be sensitive or confidential. Such datasets must be managed ethically, with respect to privacy laws and regulation. Organizations therefore should consider anonymizing data used for training AI models personally identifiable information (PII) is sufficiently protected. Moreover, breaches or misuse can lead to significant reputational damage and legal consequences. It is therefore essential to implement strict data anonymization techniques, using differential privacy, while data is stored securely and encrypted [56].

*2) Consent, Surveillance, and Proportionality*

While hunting for threat vectors, evidences, indicators of compromise, either manually or with the use of AI, oftentimes the lines between legitimate surveillance and invasion of privacy are blurred. Hence, any data collection or surveillance activities should be executed having prior consent and in accordance with legal and ethical standards. Ultimately, relevant individuals should be aware of and agree to the monitoring and data collection processes, respecting their autonomy and rights. Nonetheless, the implementation of clear consent mechanisms, regularly updating terms of service, and transparency in data collection practices can solve this challenge [57]. The scale of surveillance must also be proportionate, thus preventing overreach and potential misuse of data. Solving this challenge requires regular audits and setting clear boundaries on data collection [58].

*3) Technology Misuse*

Like any innovative tool, the AI-enhanced CTI pipeline can potentially be misused. There is a potential risk of being used for malicious purposes, for instance, spreading misinformation or running unauthorized surveillance. Organizations should implement strict access controls and behavioural monitoring to prevent misuse. NIST recently provided a thorough AI risk management framework describing this challenge and potential solutions [52].

*B. Addressing Potential Biases in AI-Enhanced CTI Models*

*1) Training Data Scrutiny*

AI models are only as good as the data they are trained on. If the training data contains biases, the AI model will likely inherit those biases. It is crucial to use training datasets that are diverse and representative to avoid unintentional biases in AI-driven insights. Biased training data will lead to skewed AI predictions, ultimately leading to unfair outcomes. To address this challenge, we need to consider diverse datasets, employing fairness-enhancing interventions, and regular bias audits. However, even with unbiased training data, algorithms themselves can introduce biases [59]. It is therefore evident that regular audits and evaluations of AI models can help to identify and rectify any inherent biases in the algorithms [60].

*2) Continuous Model Evaluation*

Continuous evaluation will provide reasonable assurance that the AI-enabled pipeline will remain relevant and unbiased as new CTI data emerges. Moreover, biases in the AI model may create loops where the model's predictions reinforce existing biases. For instance, if an AI model incorrectly flags certain types of network traffic as malicious due to biases, it may lead to increased scrutiny of similar traffic in the future, reinforcing the bias. It is therefore imperative to implement real-time evaluation metrics and periodic retraining of models [60]. If the AI-driven CTI pipeline makes a wrong decision, it is crucial to have a reliable rollback mechanism in place to restore the system to its previous working state. This can be done by creating epochs or checkpoints before carrying out any playbook actions [61]. Therefore, if a decision is found to be incorrect or harmful, the system can easily go back to a state before the action was taken, reducing the likelihood of disruptions.

*3) Systematic Bias Detection*

The AI-enhanced CTI pipeline may be prone to observer bias, which appears when the subjective predispositions of individuals involved in the AI's development or operational phases influence the selection of training data or the interpretation of the system's outputs. Such biases can inadvertently lead to misrepresentation in the AI model's both on the CTI as well as the threat detection capabilities. This will potentially result in the disproportionate identification or neglect of specific threats. To defend the precision and impartiality of the AI-enhanced CTI pipeline, it is imperative to, firstly, acknowledge and address the presence of observer bias. Moreover, training data bias, a form of availability bias, originates from the initial CTI collection phase of the pipeline.

This may occur when the training data predominantly consists of easily accessible information, or when over or undersampling leads to a training set that does not accurately represent real-world scenarios. Nonetheless, mitigations for both these biases have been proposed by Schwartz et al. (NIST) [62] and other scholars [63], such as the utilization of heterogeneous data sources and the implementation of systematic bias detection and correction mechanisms throughout the lifecycle of the AI model. Additionally, organization should consider regularly testing the AI system's performance using blind or double-blind methods, where neither the testers nor the AI system have information that might influence the outcome of the test. This would provide a method of assessing the AI system's ability to identify threats without bias.

### C. Transparency and Interpretability within the AI-Enhanced CTI Pipeline

#### 1) Explainable AI (XAI) and Stakeholder Trust

Many advanced AI models, especially deep learning models, are oftentimes seen as "black boxes" where their decision-making processes are not easily interpretable. This poses a significant challenge in CTI, where understanding the rationale behind insights is crucial for decision-making. XAI enables AI model predictions to become understandable by humans, fostering trust, and facilitating better decision-making. To address the black box issue, there is a growing stress on model explainability. SHAP (SHapley Additive exPlanations) or LIME (Local Interpretable Model-agnostic Explanations) techniques emerged as means to provide insights into how AI models arrive at their decisions. Therefore, using interpretable models, employing post-hoc explanation techniques, and visualizing model decisions is a demonstrated way forward to solve this challenge effectively [64].

Lastly, it is of utmost importance for AI-driven CTI systems to be effective, that the stakeholders must trust the insights with which they are provided. Guaranteeing transparency and interpretability is key to building and maint0aining this trust. Regular communication how the AI models work, their limitations, and the steps taken to address their accuracy can help foster trust among stakeholders.

#### 2) Feedback within the Human-AI fusion

Feedback from CTI analysts can refine AI models and therefore verify their continued relevance and accuracy. This can be achieved through implementation of an iterative refinement model [65], and by fostering a collaborative AI-human environment. Additionally, comprehensive documentation warrants that all stakeholders understand the workings, limitations, and scope of AI models, thus, maintaining detailed model logs, providing clear documentation on algorithms and training data, and confirming transparency in model updates [66].

### D. AI Model Robustness and Adversarial Attacks

Prior adopting an AI-driven CTI pipeline, a prime consideration is that the pipeline itself might become potential target for cyber adversaries. Similarly to traditional software systems that can be exploited, AI models have their own set of vulnerabilities, especially to adversarial attacks. These attacks contain feeding the model specially crafted input data designed to deceive it, leading to incorrect outputs or predictions [67]. Adversarial attacks can be categorized on a high level in two types. White-box attacks, where the attacker has complete knowledge of the AI model, the architecture and its weights, and black-box attacks, where the attacker has no knowledge of the model's internals and only has access to its inputs and outputs. Common adversarial attack techniques include adding imperceptible noise to input data, generating adversarial examples, or exploiting model transferability where an adversarial example crafted for one model affects another [68]. An adversarial attack could lead to several adverse outcomes. For instance, misclassification of benign network traffic as malicious or vice versa. Incorrect threat scoring, leading to miss prioritization of threats, or even deceptive insights that could mislead incident response teams or decision-makers [67].

To protect the AI-driven CTI pipeline against adversarial attacks, several strategies can be followed. For instance, training the AI model on adversarial examples, making it more robust against such attacks. Input filtering or normalization can detect and mitigate adversarial input data [69]. Use of collaborative models will increase robustness as an attacker would need to deceive multiple models simultaneously. Moreover, continuously updating and retraining the AI model confirms that it is equipped to manage new adversarial techniques. Organisations should also consider adopting frameworks to perform red teaming against generative AI models such as PyRIT[6], to assess and improve the security posture of the models, making them more resilient to adversarial threats and counter biases within the models. Lastly, implementing real-time monitoring to detect unusual patterns in the model's predictions can flag potential adversarial attacks.

## V. CONCLUSIONS & FUTURE RESEARCH

In this work we defined an AI-enhanced CTI processing pipeline and presented how the integration of AI into CTI processing has the potential to revolutionize the cybersecurity landscape. AI can automate, enhance, and expedite numerous CTI tasks, from data ingestion to cyber resilience verification and therefore organizations can achieve a more proactive and adaptive cybersecurity posture, staying a step ahead of evolving threats, as opposed to a reactive approach. A successful implementation would signal the beginning of an AI based end to end cyber defence. Currently, human CTI analyst and AI are bounded by strong collaborative bonds where both complement each other's strengths and counter each other's weaknesses.

---

[6] https://www.microsoft.com/en-us/security/blog/2024/02/22/announcing-microsofts-open-automation-framework-to-red-team-generative-ai-systems/

However, the integration of AI into CTI brings implementation challenges, ethical considerations, potential biases, and the need for transparency and interpretability that require attention. Nonetheless, with a balanced approach that combines the strengths of AI with human expertise, these challenges can be addressed effectively.

With this work we set the foundational framework for an AI-enhanced CTI processing pipeline, and we identify several research routes. Some potential directions for future research may be advanced AI Models. As AI continues to evolve, exploring newer models and architectures tailored for CTI tasks can yield even more accurate and efficient results. Another angle is the ethical AI in cybersecurity. A thorough research into the ethical implications of AI in cybersecurity, developing guidelines and best practices for responsible deployment.

Moreover, on the Human-AI collaboration, further research on optimizing the collaboration between AI systems and human analysts, can demonstrate how each complements the other's strength. Researching the integration of more diverse and unconventional data sources into the CTI pipeline, such as social media could potentially allow for a more accurate prediction of the threat actor next attacks. Another research route is to study ways to mitigate the biases in the pipeline and explore ways to make risk decisions about which runbacks could be automated and which cannot. Lastly, a more technical angle would be to investigate the feasibility and methodologies for real-time threat intelligence processing using AI, enabling instantaneous response to emerging threats while also achieving automated compliance to security policies, which is our next focus.

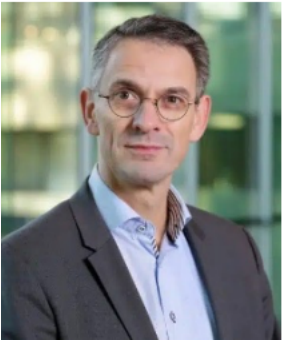

Martijn Dekker received the M.Sc. degree in mathematics at the University Utrecht in 1993 and the Ph.D. degree in mathematics at the University of Amsterdam in 1997. He has been an associate professor at the TIAS business school for Business and Society from 2012 to 2020. Since 2020 he is visiting professor Information Security at the University of Amsterdam / Amsterdam Business School. He is also the Chief Information Security Officer (CISO) at ABN AMRO.

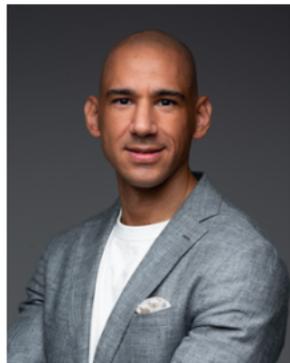

Lampis Alevizos received his M.Sc. degree in cybersecurity and his Ph.D. degree in Computer Science from the University of Central Lancashire (UCLan), with research focused on the convergence of ZTA, blockchain, and DLT with cybersecurity. Lampis holds and actively maintains several industry certifications, including CISSP, CCSP, CCSK, CISA, CISM, among others. He is currently the Head of Cyber Defense Innovation at Volvo Group.

The paper reflects the author's own personal views, not necessarily those of ABN AMRO or Volvo Group.